# Shock Synthesis of Organic Molecules by Meteoroids in the Atmosphere of Titan

Erin E. Flowers[1,3] and Christopher F. Chyba[1,2]

[1] Department of Astrophysical Sciences, Princeton University, 4 Ivy Lane, Princeton, NJ 08540, USA; eflowers@princeton.edu
[2] School of Public and International Affairs, Princeton University, 20 Prospect Lane, Princeton, NJ 08540, USA

## Abstract

Thermochemical modeling and shock-tube experiments show that shocks applied to $N_2/CH_4$ gas mixtures can synthesize organic molecules. Sufficiently large, hypersonic meteoroids entering the atmosphere of Saturn's moon Titan should therefore drive organic chemistry. To do so meteoroids must be sufficiently large compared to the atmospheric mean free path at a given altitude to generate shocks, and deposit enough energy per path length to produce temperatures high enough to excite and dissociate the relevant molecules. The Cassini spacecraft imaged multiple meteoroid impacts on Saturn's rings, allowing for the first time an empirical estimate to be made of the flux and size–frequency distributions of meteoroids in the millimeter-to-meter size range. We combine these results with an atmospheric entry model and thermochemical and experimental shock production efficiencies for $N_2/CH_4$ atmospheres and calculate the shock production rates for HCN, $C_2H_2$, and $C_2H_4$ as well as the resulting $H_2$ generation. We find that meteoroids may be producing these molecules at as much as ∼1% the production rate of photochemistry driven by UV photons, and may be depositing more energy than magnetospheric ions and 90–100 nm UV photons. Moreover, these meteoroids produce these organic molecules hundreds of kilometers lower in Titan's atmosphere than the relevant UV photons and magnetospheric ions penetrate, with peak production occurring between 200 and 500 km altitudes, i.e., at the observed haze layer. Meteoroid-driven shock generation of molecules may therefore be crucial to understanding Titan's atmospheric chemistry.

*Unified Astronomy Thesaurus concepts:* Titan (2186); Natural satellite atmospheres (2214)

## 1. Introduction

Saturn's moon Titan is the only planetary satellite in our solar system with a dense (∼1.5 bar) atmosphere, which has an overall composition of 94.2% $N_2$, 5.65% $CH_4$, 0.1% $H_2$, and smaller amounts of nitriles, hydrocarbons, and other organics (Strobel 2010; Hörst 2017). This atmosphere provides an arena for abundant organic chemistry, thought to be driven primarily by solar ultraviolet (UV) and charged particle radiation (Sagan & Thompson 1984; Yung et al. 1984; Krasnopolsky 2009; Snowden & Yelle 2014). In the upper atmosphere, molecular nitrogen and methane are dissociated by UV and charged magnetospheric particles, resulting in various C-H-N species. These products then participate in a wide variety of chemical reactions with each other and the ambient neutral and ionized species (such as magnetospheric $O^+$; Hartle et al. 2006) to produce additional hydrocarbons, neutral atomic species, and other C-H-N-O species as described in detail by, for example, Sagan & Thompson (1984), Yung et al. (1984), Cabane & Chassefière (1995), Krasnopolsky (2009), and Snowden & Yelle (2014). Neutral hydrogen species (H and $H_2$) in particular are produced through the dissociation of methane and ambient hydrocarbons, but are quickly lost from the atmosphere due to Titan's low gravity. Various studies show that neutral hydrogen production peaks below 1000 km, with atomic hydrogen production peaking at roughly 800 km (Lebonnois et al. 2003) and molecular hydrogen production peaking at roughly 550 km (Krasnopolsky 2009).

Titan's atmosphere is also subject to a constant bombardment of dust particles and meteoroids. Many families of particles contribute to this flux, including interplanetary dust particles (IDPs) of various origins as well as material from Saturn's extensive ring system. Sources of interplanetary material include Jupiter-family comets (JFCs), Kuiper Belt objects (KBOs), Halley-type comets (HTCs), and Oort Cloud comets (OCCs; Landgraf et al. 2002). Cassini–Huygens spacecraft measurements show that Saturn's E ring extends out to Titan and beyond (Srama et al. 2006, 2011), providing a population of dust impactors on Titan that originate with Saturn's active moon Enceladus.

Particles falling into Titan's atmosphere are an important source of material that can participate in chemical reactions (English et al. 1996; Molina-Cuberos et al. 2001). Sufficiently large particles moving at hypersonic speeds will also generate atmospheric shocks (Lin 1954; Revelle 1976; Silber & Brown 2014; Silber et al. 2017, 2018). Shocks in an $N_2/CH_4$ atmosphere have long been known to be an especially efficient source of organic synthesis (e.g., Rao 1966; Rao et al. 1967; Bar-Nun & Shaviv 1975; Borucki et al. 1988; Scattergood et al. 1989; Cabane & Chassefière 1995; Hörst et al. 2018), with experimental yields (molecules produced per joule of input energy) higher than those for UV or charged particles for some species (Scattergood et al. 1989; Chyba & Sagan 1992). However, particles whose spherical radius ($R$) is sufficiently smaller than the atmospheric mean free path ($\lambda$) at a given altitude in the atmosphere will not generate shocks at that altitude, because the atmosphere cannot be compressed by the particle (Silber et al. 2018). (This criterion will be rendered more specifically below.) For specificity we will label as "meteoroids" those particles large enough to generate shocks at some point during their atmospheric entry, ablation, and deceleration, and those that are too small to generate shocks as "dust" particles.

---

[3] NSF Graduate Research Fellow.

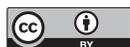

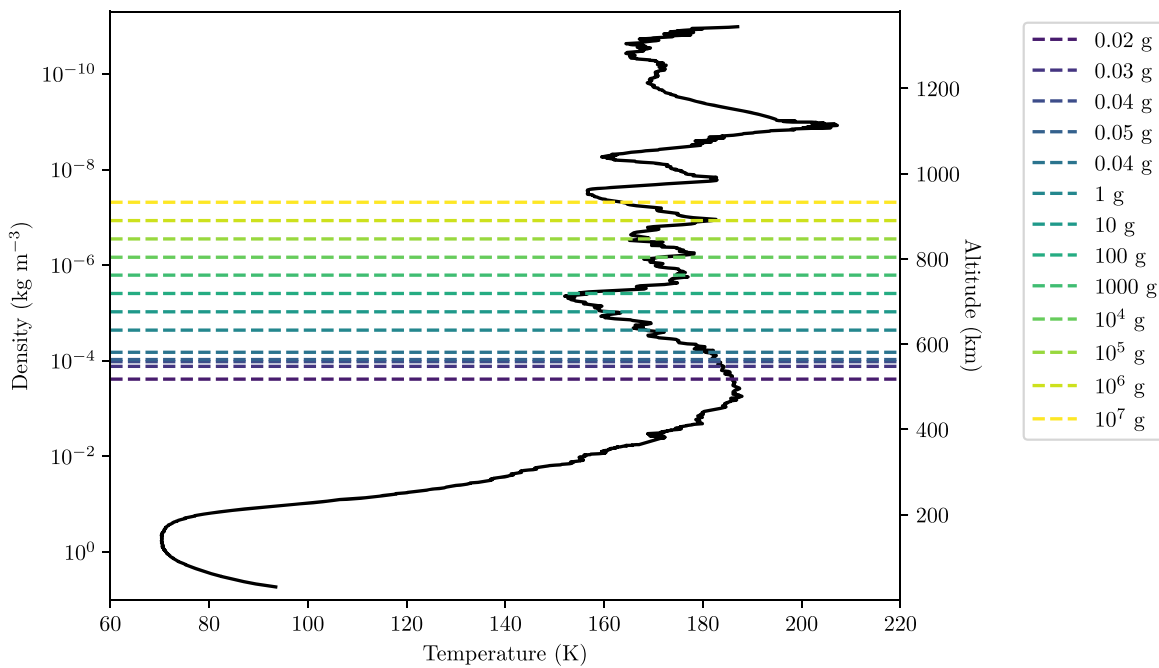

**Figure 1.** Temperature–density–altitude profile of Titan's atmosphere as measured by HASI. The dashed lines note where the particles in our study begin to experience shock formation in the atmosphere; the criterion for the shocking condition will be described in Section 3.

Meteoroid-generated shock waves of sufficient energy have previously been recognized as a potential energy source for organic synthesis on Titan (Scattergood et al. 1989), but the importance of the effect could not be reliably quantified, because only the microscopic dust flux at Saturn had been measured. However, the Cassini spacecraft imaged the results of meteoroid impacts on Saturn's rings, allowing for the first time an observational estimate to be made of the flux and size–frequency distributions of meteoroids at Saturn, and therefore at Titan (Tiscareno et al. 2013). Here, we combine these observationally derived meteoroid flux models and Huygens atmospheric data with yields for organic shock synthesis in $N_2/CH_4$ gas mixtures from theoretical calculations and shock-tube experiments to predict the quantity of organics expected to be produced in Titan's atmosphere by meteoroid infall.

In Figure 1 we show the temperature–density profile as measured by the Huygens Atmospheric Structure Instrument (HASI; Fulchignoni et al. 2005) that we use in our model. The horizontal lines correspond to where in the atmosphere the particles begin to enter the flow regime favorable to shock formation. This is the altitude where they meet the threshold value for the Knudsen number, which we will discuss in more detail in Section 3.

In Section 2 we present flux models for microscopic dust and meteoroids at Titan, calculating impact velocities and the effects of gravitational focusing to both Saturn's and Titan's gravitational attraction. We then calculate the total mass, and therefore shock energy, available from the meteoroids to drive organic synthesis. This requires determining the smallest meteoroid particle size capable of generating a shock in Titan's atmosphere, and this in turn requires the use of an atmospheric entry model since $\lambda$ varies with altitude. We present this model in Section 3. In Section 4 we review relevant experimental and theoretical results for $N_2/CH_4$ shock synthesis. In Section 5, we calculate the overall shock production of several organic molecules in Titan's atmosphere. Finally, in Section 6, we compare these results with those from a standard photochemical model, discuss uncertainties, and draw conclusions. While there are significant uncertainties (quantified below), organic shock production from meteoroids appears to be significant enough that Titan atmospheric chemistry models must henceforth take these effects into account.

## 2. Impact Fluxes and Velocities

There are a number of models for the dust flux at Saturn as a function of particle mass. We summarize some of the most prevalent of these in Figure 2. Grün et al. (1985), Divine (1993), Poppe & Horányi (2012), and Poppe (2016) predict fluxes for particles between 0.5 and 100 $\mu$m, originating from JFCs, HTCs, the Kuiper Belt, and the Oort Cloud, and while these microscopic particles have the largest flux into the Saturnian system, we will see in Section 3 that they are too small to drive shock chemistry in Titan's atmosphere.

However, Tiscareno et al. (2013) report observations taken by Cassini's Imaging Science Subsystem (Porco et al. 2004) of ejecta clouds produced by impacts on Saturn's A, B, and C rings that allow them to estimate the flux ($\Phi$) as a function of radius ($R$) for meteoroids in the centimeter-to-meter range. Meteoroids in this size range are easily large enough to generate shocks in Titan's atmosphere. Tiscareno et al. (2013)'s flux estimates (their value for $\Phi$) do not account for two effects that are important for our modeling: gravitational focusing by Saturn (by which Saturn's gravity both increases infalling particle velocities and increases their flux by drawing them in toward Saturn), and the two-dimensional nature of the target presented by Saturn's rings. Their flux estimates must therefore be adjusted by some velocity-dependent factor we write as $f_\infty$ that accounts for these effects. The gravitational focusing factor at some point in space a distance ($r$) from the center of Saturn is

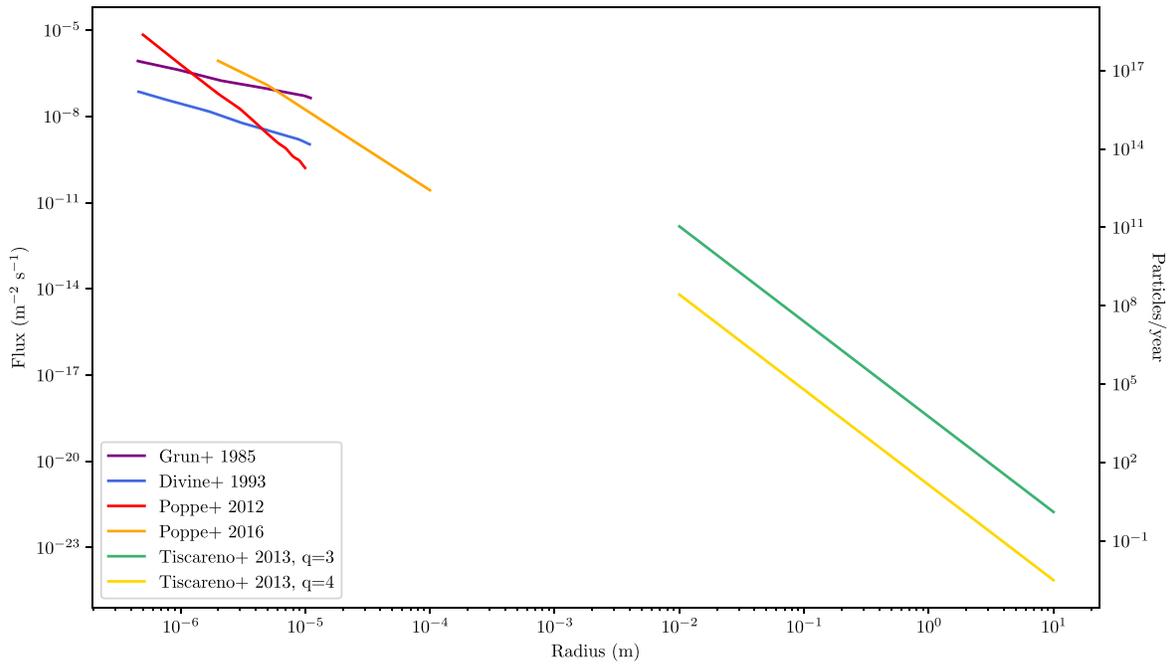

**Figure 2.** Comparison of model fluxes for dust particles and meteoroids in the outer solar system at the heliocentric distance 9.5 au. The Poppe & Horányi (2012) and Poppe (2016) models were constrained by data from the New Horizon's Student Dust Counter, while those of Tiscareno et al. (2013) were constrained by observations of ring impacts seen by the Cassini orbiter. As described in the text, we have adjusted the Tiscareno et al. (2013) models to correct for the gravitational focusing effects of Saturn, and have changed their model to assume a density of 3400 kg m$^{-3}$ (see text), rather than their choice of 1000 kg m$^{-3}$.

given by Colwell (1994):

$$G(r) = 1 + \frac{1}{2}\left(\frac{v_{Sesc}(r)}{v_\infty}\right)^2, \quad (1)$$

where:

$$v_{Sesc}(r) = (2GM_S/r)^{1/2}, \quad (2)$$

is Saturn's escape velocity at $r$, with $G = 6.67 \times 10^{-11}$ m$^3$ kg$^{-1}$ s$^{-2}$ the gravitational constant, $M_S = 5.68 \times 10^{26}$ kg the mass of Saturn, and $v_\infty$ the meteoroid's velocity at "infinity"—say at Saturn's Hill radius. Tiscareno et al. (2013)'s results depend most strongly on B- and C-ring impacts, so we take $r = 92{,}000$ km, or 1.6 $R_S$ (Saturnian radii), corresponding to the boundary between the B and C rings. For this $r$, we have $v_{Sesc}(r) = 28.7$ km s$^{-1}$ from Equation (2).

Poppe (2016) finds that the IDP mass flux at Saturn's Hill radius is dominated by the OCC population, consistent with Grün et al. (1985)'s earlier conclusion. Poppe calculates the median velocity distribution for this family of grains (with radii of approximately 0.5–100 $\mu$m), and finds it to be 16 km s$^{-1}$, again before any acceleration due to Saturn's gravity. Cuzzi & Durisen (1990) argue that meteoroids impacting Saturn's rings are also primarily Oort Cloud–originating objects. Taking $v_\infty = 16$ km s$^{-1}$ for OCC-originating meteoroids gives $G$ ($r = 1.6 \ R_S$) = 2.6 by Equation (1). (We note that IDPs experience size-dependent nongravitational forces such as radiation pressure and Poynting–Robertson drag that are insignificant for larger meteoroids; we ignore these complications here.) The two-dimensional nature of Saturn's rings reduces the extrapolation of the flux to the Hill radius by another factor of two (Tiscareno et al. 2013), so that we take $f_{esc} = 1/(2 \times 2.6) = 0.19$ as the multiplicative factor needed to extrapolate from a flux at the B/C ring boundary to "infinity."

Our plot of the Tiscareno et al. (2013) data in Figure 2 uses this value for $f_\infty$.

To extrapolate $\Phi$ to Titan, we must then account for the gravitational focusing due to Saturn's escape velocity at Titan's distance of 20.3 $R_S$. At this distance, $v_{Sesc}(r) = 7.9$ km s$^{-1}$ and $G$ ($r = 20.3 \ R_S$) = 1.1. Titan's escape velocity $v_{Tesc}$ at the top of its atmosphere, which in our simulations we will take to lie at an altitude of 1500 km, is 2.10 km s$^{-1}$. This increases $G(r)$ by only about 1%. Extrapolating $\Phi$ at infinity to Titan therefore requires multiplying by a factor we write as $f_T = 1.1$. A typical meteoroid impact velocity $v_i$ at 1500 km above Titan's surface is given by:

$$v_i^2 = v_\infty^2 + v_{Sesc}^2 + v_{Tesc}^2, \quad (3)$$

which gives $v_i = 18$ km s$^{-1}$. As a consistency check, we note that the impact velocity at, say, Saturn's B/C ring boundary resulting from this approach is 33 km s$^{-1}$, consistent with the velocities found by Cuzzi & Durisen (1990) in their work on Saturnian ring impacts from Oort Cloud meteoroids.

Tiscareno et al. (2013) adopt a power-law distribution for the ejecta produced by meteoroid impact with Saturn's rings, viz $n(s) = n_0 s^{-q}$, where $n(s)ds$ is the number of ejecta particles with radius between $s$ and $s + ds$, and $3 < q < 4$. For the two end-member cases $q = 3$ and $q = 4$, their regression over their data ultimately yields two possible equations for the meteoroid flux $\Phi$:

$$\log \Phi_3 = -19.476 - 3.643 \log R, \quad (4)$$

and:

$$\log \Phi_4 = -22.086 - 3.643 \log R, \quad (5)$$

where $R$ is the (assumed spherical) meteoroid radius in meters and $\Phi_q$ has units of m$^{-2}$ s$^{-1}$. These are cumulative fluxes, meaning that they give the number flux of all meteoroids with radii $\geqslant R$. Note that the corresponding equations in Tiscareno et al. ([2013](#)) are in error due to a calculation mistake. Equations ([4](#)) and ([5](#)) above are the correct expressions (Tiscareno, personal communication, 2022). Although not stated explicitly in their paper, Tiscareno et al. ([2013](#)) assumed a spherical meteoroid density $\delta = 1000\ \mathrm{kg\ m^{-3}}$ in deriving these equations (Tiscareno, personal communication, 2022). Equations ([4](#)) and [5](#) can also be written as:

$$\Phi_q = a_q R^b, \qquad (6)$$

where $a_3 = 3.34 \times 10^{-20}\ \mathrm{m^{-2-b}\ s^{-1}}$, $a_4 = 8.20 \times 10^{-23}\ \mathrm{m^{-2-b}}$ $\mathrm{s^{-1}}$, and $b = -3.643$. These are all constants that are determined from fits to the data.

We note that recent models for microscopic dust at Saturn assume a higher density, with $\delta$ between 2500 and 4000 $\mathrm{kg\ m^{-3}}$ (e.g., Poppe [2016](#)), and that this has been corroborated by some IDP and meteorite density determinations (e.g., Flynn [2004](#); Macke et al. [2011](#)). We therefore change the impactor density $\delta$ implicit in Equations ([4](#)) and ([5](#)) from 1000 $\mathrm{kg\ m^{-3}}$ to 3400 $\mathrm{kg}$ $\mathrm{m^{-3}}$, a value that is also used in several recent models, particularly those that cite Ip ([1990](#)) for parameters. Because $R \propto \delta^{-1/3}$, this is equivalent to multiplying $R$ by a factor 0.6650. Equations ([4](#)) and ([5](#)) then become:

$$\log \Phi_3 = -18.831 - 3.643 \log R, \qquad (7)$$

and:

$$\log \Phi_4 = -21.441 - 3.643 \log R, \qquad (8)$$

and these are the equations that we have used for the meteoroid flux in Figure [2](#). There is a gap in the models shown in Figure [2](#) between $\sim 10^{-7}$ and $\sim 10^{-4}$ g, due to the lack of observational evidence for particles in this mass range—they are too large to be detected by spacecraft dust impact instruments, but too small (so far) to be probed by optical observations.

Following Equations ([7](#)) and ([8](#)), the coefficients in Equation ([6](#)) for the cumulative flux become $a_3 = 1.48 \times 10^{-19}\ \mathrm{m^{-2-b}\ s^{-1}}$, $a_4 = 3.63 \times 10^{-22}\ \mathrm{m^{-2-b}\ s^{-1}}$, and again $b = -3.643$. The differential flux (the flux per particle radius) is given by:

$$\partial \Phi_q / \partial R = b a_q R^{b-1}, \qquad (9)$$

where $q = 3$ or 4. We can then calculate the total mass flux of meteoroids (assumed spherical) in the range of radii from $R_2$ to $R_1$ as:

$$M_q = f_T f_\infty \int_{R_2}^{R_1} \frac{4}{3} \pi \delta R^3 \left( \frac{\partial \Phi_q}{\partial R} \right) dR, \qquad (10)$$

where $\Phi_q$ is given by either Equation ([7](#)) or ([8](#)), and $M_q$ has units of mass flux, $\mathrm{kg\ m^{-2}\ s^{-1}}$. By Equation ([9](#)) this gives:

$$M_q = f_T f_\infty \frac{4}{3} \pi \delta \left( \frac{b}{b+3} \right) a_q (R_1^{b+3} - R_2^{b+3}). \qquad (11)$$

By Equation ([11](#)), if $R_2 \gg R_1$, $M_q$ is dominated by the choice of $R_1$—the smallest meteoroid included in the calculation.

In principle one should also integrate over an intrinsic velocity distribution about $v_i$, but this distribution is unknown for centimeter-to-meter-scale meteoroids at Saturn. Titan orbits Saturn with an orbital velocity of $v_{Torb} = 5.6\ \mathrm{km\ s^{-1}}$, so that impact velocities vary sinusoidally from its leading to trailing edge like $v_i + v_{Torb} \cos \psi$, where $\psi$ is 0 at Titan's leading point and $\pi$ at Titan's trailing point. Obviously this does not change the average impact velocity with Titan (though it does represent a significant hemispheric variation); we defer examination of the effects of this variation to a subsequent investigation.

## 3. Atmospheric Entry and Shock Generation

What is the smallest meteoroid that will generate a shock in Titan's atmosphere? This depends on the Knudsen number ($Kn$), defined as the ratio of the atmosphere's mean free path ($\lambda$) to the particle radius ($R$):

$$Kn = \frac{\lambda}{R}. \qquad (12)$$

When $Kn$ is sufficiently large (in the free molecular flow regime) there can be no shock. Intuitively, one might expect that shocks should only set in for $Kn \ll 1$. But while that may hold for reentry vehicles, it does not hold for meteoroids, for two reasons (Silber et al. [2018](#); Moreno-Ibáñez et al. [2018](#)). One is that the ablating meteoroid forms a vapor cap whose dimensions can be up to two orders of magnitude larger than the original meteoroid itself. The other is that the mean free path around the ablating meteoroid is smaller than the mean free path in the atmosphere. And in fact meteor shocks in Earth's atmosphere set in at much higher altitudes (corresponding to much larger values of $Kn$) than would be the case for reentry vehicles. Observationally, meteors with $R \gtrsim 2$ mm create shocks in Earth's upper atmosphere, at altitudes of around 90 km and below.

At 90 km, Earth's atmospheric number density is $n = 7.12 \times 10^{19}\ \mathrm{m^{-3}}$, and $\lambda = 2.37$ cm (US Standard Atmosphere [1976](#)), corresponding to $Kn = 12$ in Equation ([12](#)). The mean free path is $\lambda = (n\sigma)^{-1}$, where $\sigma$ is the cross section for collisions for the atmospheric molecules. Earth's atmosphere is dominated by $N_2$ and $O_2$; for $N_2$, $\sigma = 0.43\ \mathrm{nm^2}$ and for $O_2$, $\sigma = 0.40\ \mathrm{nm^2}$ (Atkins [1986](#)). Titan's atmosphere is dominated by $N_2$ with about 5% $CH_4$, for which $\sigma = 0.46\ \mathrm{nm^2}$ (Atkins [1986](#)), so that $\sigma$ in Titan's atmosphere may be taken to be the same to within a few percent as that of Earth's. We can therefore take shock production in Titan's atmosphere to occur for meteoroids of radius $R$ that penetrate deeply enough to reach a value of $n$ corresponding to a $\lambda$ that gives the same $Kn$ threshold value for shock creation as in Earth's atmosphere, viz $Kn = 12$.

To determine this $R$, we use an RK4 method to solve the equations of meteoroid motion and ablation from Bronshten ([1983](#)) and Campbell-Brown & Koschny ([2004](#)), converting from a time step to an altitude step and using data from the Huygens lander's HASI instrument for atmospheric density at a given altitude (Fulchignoni et al. [2005](#)). We solve from the top of the atmosphere (defined here to be at $z = 1400$ km) to the surface.

The energy equation is (Campbell-Brown & Koschny [2004](#)):

$$\frac{dT}{dz} = \frac{1}{Cmv \cos \theta} \left[ \frac{\Lambda \rho v^3}{2} A \left( \frac{m}{\delta} \right)^{2/3} \right.$$
$$\left. - 4\sigma_{SB} \epsilon (T^4 - T_a^4) A \left( \frac{m}{\delta} \right)^{2/3} - L \frac{dm}{dz} \right], \qquad (13)$$

where $C$ is the specific heat, $m$ is the mass of the particle, $v$ is its velocity, $\theta$ is the angle of its trajectory with respect to the vertical, $\Lambda$ is the heat transfer coefficient, $\rho$ is the atmospheric density at the altitude step, $\delta$ is the meteoroid density, $\sigma_{SB}$ is the Stefan–Boltzmann constant, $\epsilon$ is the emissivity of the particle, $T$

| Name | Variable | Value | Reference |
|------|----------|-------|-----------|
| Specific heat | $C$ | $9 \times 10^6$ erg g$^{-1}$ K$^{-1}$ | Campbell-Brown & Koschny (2004) |
| Heat transfer coefficient | $\Lambda$ | 0.5 | Campbell-Brown & Koschny (2004) |
| Meteoroid material density | $\delta$ | 3.4 g cm$^{-3}$ | Ip (1990) |
| Emissivity | $\epsilon$ | 0.9 | Campbell-Brown & Koschny (2004) |
| Shape factor | $A$ | 1.2 | Bronshten (1983) |
| Heat of ablation | $L$ | $8.1 \times 10^{10}$ erg g$^{-1}$ | Ip (1990) |
| Drag coefficient | $\Gamma$ | 1 | Campbell-Brown & Koschny (2004) |
| Lift coefficient | $C_L$ | $10^{-3}$ | Chyba et al. (1993) |

is its temperature, $T_a$ is the atmosphere's temperature at the altitude step, $A$ is the shape factor, and $L$ is the heat of ablation. The velocity equation is:

$$\frac{dv}{dz} = -\frac{1}{\cos\theta}\frac{\Gamma\rho v}{m}A\left(\frac{m}{\delta}\right)^{2/3} + \frac{g}{v}, \qquad (14)$$

with:

$$g = g_T\left(\frac{R_T}{R_T + z}\right)^2, \qquad (15)$$

where $g_T = -1.352$ m s$^{-2}$ is the gravitational acceleration of Titan at surface level, $R_T$ is the radius of Titan, and $\Gamma$ is the drag coefficient. Finally, for the ablation equation we take (Bronshten 1983):

$$\frac{dm}{dz} = -\frac{1}{\cos\theta}\frac{A\Lambda}{2L}\left(\frac{m}{\delta}\right)^{2/3}\rho v^2. \qquad (16)$$

Particles entering the atmosphere can also lose mass due to sputtering, which on Earth may set in at altitudes much higher than those at which ablation modeled by Equation (16) becomes important. (Sputtering is due to direct collisions of atmospheric molecules with the surface of the meteoroid, thereby dislodging surface material.) But this process appears to be important only for incident particle velocities above about 30 km s$^{-1}$. Above this velocity, some incident particles (depending on mass and density) may sputter away $\sim$10%–20% of their initial mass (Hill et al. 2004; Popova et al. 2004). Given the average meteoroid impact velocity of 18 km s$^{-1}$ at Titan found here, we ignore the effects of sputtering in this discussion.

Finally we also model how $\theta$ changes using the equation from Chyba et al. (1993):

$$\frac{d\theta}{dz} = \left(\frac{1}{v\cos\theta}\right)\left(-\frac{g\sin\theta}{v} + \frac{C_L\rho S_A v}{2m} + \frac{v\sin\theta}{R_T + z}\right). \qquad (17)$$

The values of the constants used in our atmospheric entry simulations are shown in Table 1, along with the references for these choices. In the free molecular flow regime, we expect $\Lambda \approx 1$ (Grebowsky 1981), and this is the choice commonly made in the literature when modeling the atmospheric entry of microscopic dust. For objects large enough to generate atmospheric shock waves (meteoroids), however, this number is too high, and observations suggest $\Lambda = 0.5$ is a more appropriate choice (Bronshten 1983; Campbell-Brown & Koschny 2004). Because we choose $\delta = 3400$ kg m$^{-3}$ for meteoroid density, for consistency we use parameters appropriate for carbonaceous meteoroids.

Figure 3 shows the results of our atmospheric entry simulations for meteoroids of different masses. Larger particles penetrate more deeply into the atmosphere before ablating away. Figure 3 also shows the $Kn = 12$ threshold; particles penetrating below this line generate shocks. We see that shocks will be generated by meteoroids with $m \geqslant 0.02$ g or $R \geqslant 1.1$ mm. To within a factor of two, this is in agreement with the size threshold for shock generation at Earth (Silber et al. 2018), despite Titan's more extended atmosphere.

## 4. Meteoroid Shock Synthesis of Organic Molecules

Atmospheric shocks produced by meteoroids entering Earth's atmosphere have been extensively modeled (Lin 1954; Revelle 1976; Silber 2014; Silber et al. 2017, 2018). The meteoroid loses its kinetic energy $E = (1/2)mv^2$ in the atmosphere by both deceleration and ablation; this loss per path length is just:

$$\frac{dE}{dz} = mv\frac{dv}{dz} + \frac{v^2}{2}\frac{dm}{dz}, \qquad (18)$$

where we have used $d/dt = vd/dz$. The $dv/dz$ deceleration term in Equation (18) generates a bow shock wave at the front of the meteoroid; the $dm/dz$ term leads to a cylindrical shock wave that trails behind the object (Silber et al. 2017). Using Equations (14) and (16), and ignoring the gravitational acceleration term, which is unimportant compared to the drag term at the altitudes in the atmosphere where our meteoroids deposit the bulk of their energy and end their trajectories, the comparative contribution of ablation and deceleration to $E$ is:

$$\frac{\frac{v^2}{2}\frac{dm}{dz}}{mv\frac{dv}{dz}} = \frac{\Lambda v^2}{4\Gamma L}. \qquad (19)$$

For the parameter values in Table 1 this ratio equals 5 at our initial velocity of 18 km s$^{-1}$ and decreases as the meteoroid decelerates, equaling 1 at a velocity of 8 km s$^{-1}$. That is, while a significant amount of the meteoroid's kinetic energy feeds the creation of the bow shock, the majority of its energy feeds the cylindrical shock until the meteoroid has lost $\sim$80% of its initial kinetic energy.

The effect of the shock passing through a given parcel of the atmosphere is to heat that parcel to high temperatures, driving chemical reactions by exciting and dissociating (depending on the duration and magnitude of the temperature spike caused by the shock) some of the atmospheric molecules. After the passing of the shock, these excited molecules and molecular fragments continue to recombine to form new species until the temperature falls to below a particular species' "freeze-out" temperature $T_F$. This is the temperature at which the chemical lifetime of that species exceeds the characteristic cooling time of the shocked atmosphere. The resulting concentrations of particular synthesized species are "frozen in" and persist even as temperatures fall far below $T_F$ (Chameides 1979; Chameides & Walker 1981).

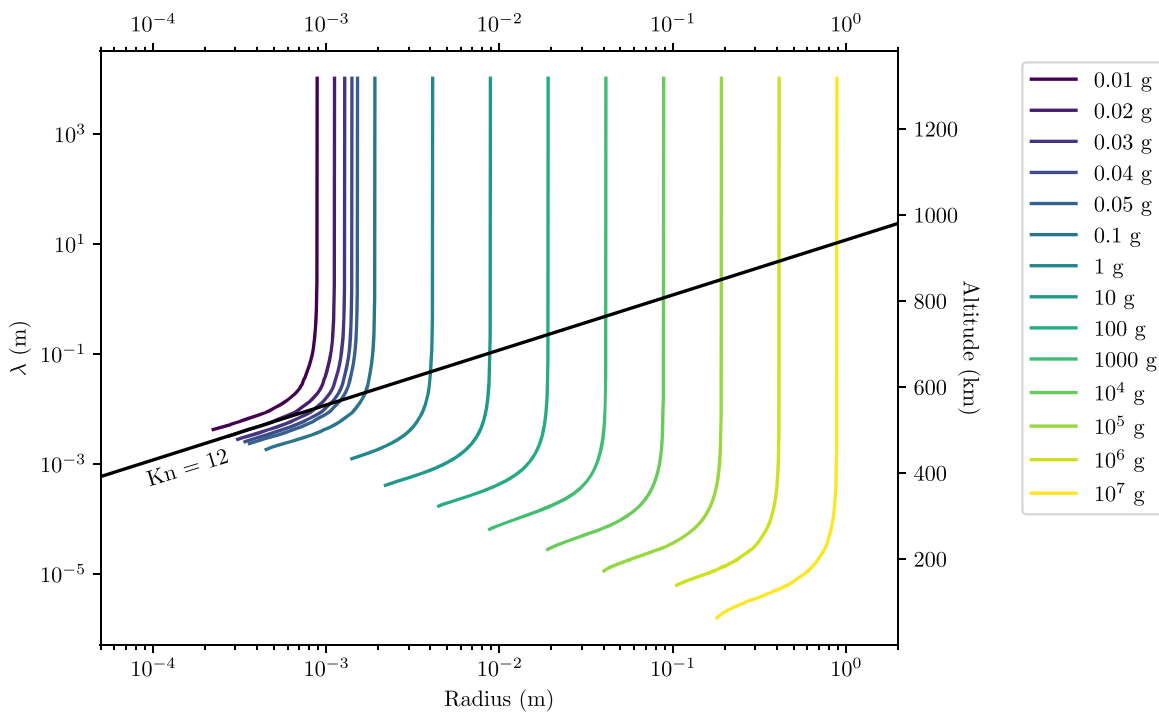

**Figure 3.** Particle mass and radius vs. altitude and mean free path $\lambda$ for ablating meteoroids in Titan's atmosphere. Meteoroids whose trajectories reach below the black $Kn = 12$ line (corresponding to meteoroids with masses greater than 0.02 g) generate shocks that may lead to organic synthesis.

A number of authors have modeled this process for Earth's contemporary $N_2/O_2$ atmosphere (e.g., Chameides 1979; Silber et al. 2017), or its early, possibly more reducing atmosphere (Chameides & Walker 1981; Ferus et al. 2017). Here we consider analogous work for $N_2/CH_4$ atmospheres appropriate for modeling Titan. We use the $N_2/CH_4$ atmosphere thermochemical, hydrodynamic, cylindrical shock chemistry calculations of Chameides & Walker (1981) that are built on the cylindrical shock physics calculations of Lin (1954). We find good agreement between the results of their calculations and results from shock-tube experiments for $N_2/CH_4$ (Rao 1966; Rao et al. 1967; Bar-Nun & Shaviv 1975) gas mixtures, with support from more recent shock-tube experiments examining $N_2/CH$ reactions (Dean et al. 1991; Lindackers et al. 1991). Finally, we compare these results with those of laser-induced plasma (LIP) experiments with $N_2/CH_4$ gas mixtures (Borucki et al. 1988; Scattergood et al. 1989; Ferus et al. 2017).

Silber et al. (2017) discuss the shock chemistry that results from meteoroids entering Earth's atmosphere, treating the deposition of energy behind the meteoroid as an exploding cylindrical line source (Lin 1954). They state that energy deposited per unit length, $E_0$, may reach $\sim 10^3$ J m$^{-1}$, leading to temperatures behind the resulting cylindrical shock wave as high as 6000 K. (Temperatures in the vapor cap at the front of the meteoroid will be significantly higher, likely in excess of 10,000 K; Silber et al. 2017; Anderson 2019; but the bulk of the object's kinetic energy is released in the trailing cylindrical shock wave, so the focus is on that phenomenon.) Molecular oxygen ($O_2$) has a dissociation energy of 5.12 eV molecule$^{-1}$ (118 kcal mole$^{-1}$), and begins to dissociate at shock temperatures as low as 2000 K, with nearly all of it dissociated by 4000 K. Molecular nitrogen

($N_2$) has a much higher dissociation energy of 9.76 eV molecule$^{-1}$ (225 kcal mole$^{-1}$), and does not dissociate until reaching temperatures above 4000 K. Nevertheless nitric oxide (NO) begins to be produced at temperatures as low as 2000 K (Anderson 2019, Figure 11.12). This is because there are at least two separate paths to the production of NO. Nitric oxide may be produced by a path in which $N_2$ is first dissociated, followed by the exothermic reaction $N + O_2 \rightarrow NO + O$. But it may also result from a Zel'dovich mechanism in which a more easily produced oxygen atom O reacts with a vibrationally excited molecular nitrogen molecule, $N_2^*$, according to O $+ N_2^* \rightarrow NO + N$. The activation energy for this reaction is only 3 eV molecule$^{-1}$ (69 kcal mole$^{-1}$), lower than either the $O_2$ or $N_2$ dissociation energy (Fridman 2008). The vibrational excitation $N_2 \rightarrow N_2^*$ of the $N_2$ molecule is itself a resonance process driven by the impact of electrons with energies of 1.7 $-3.5$ eV (Fridman 2008).

The Zel'dovich mechanism for NO production on Earth points to the explanation for why the production of hydrogen cyanide (HCN) in shocked $N_2/CH_4$ atmospheres begins at temperatures as low as 2500 K. Rao reports data from twenty-one shock-tube experiments with $N_2/CH_4$ gas mixtures in a 1:1 ratio diluted in Ar for shock temperatures between 2000 and 5750 K (Rao 1966; Rao et al. 1967). Generation of acetylene ($C_2H_2$), ethane ($C_2H_4$), and carbon "soot" is observed at 2000 K and higher temperatures; HCN production begins by 2500 K and increases toward higher temperatures, with a corresponding drop in $C_2H_2$ production as more C is incorporated into HCN. We therefore expect this "low-temperature" production of HCN to occur on Titan as meteoroids shock its atmosphere.

Rao (1966) and Rao et al. (1967) find the dissociation energy for methane in the equation $CH_4 \rightarrow CH_2 + H_2$ to be 81 kcal mole$^{-1}$, less than the dissociation energy of $O_2$. Production of HCN is unlikely to result from $N_2$ dissociation unless temperatures reach above 4000 K. Rao (1966; Rao et al.

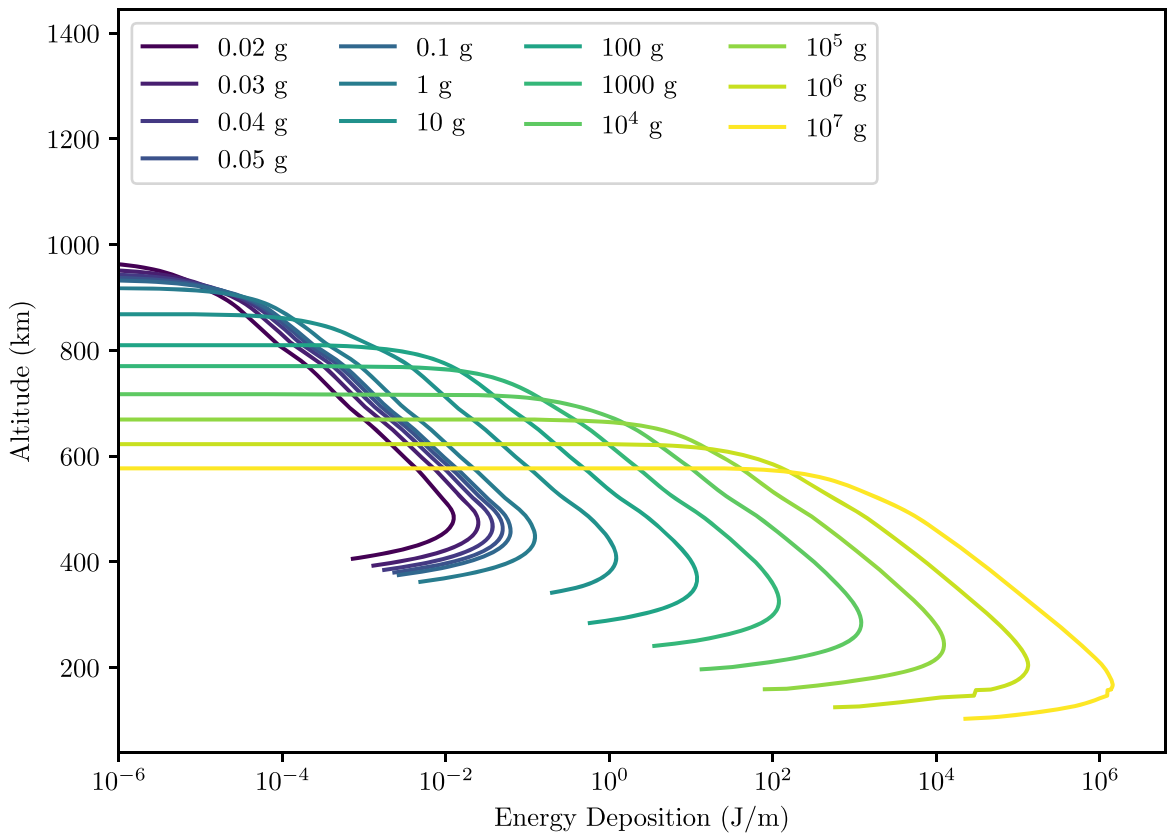

**Figure 4.** Energy deposition $E_0$ in J m$^{-1}$ for individual meteoroids of a given mass descending through Titan's atmosphere. The inversion at lower altitudes occurs as the particles begin to both decelerate and lose mass, resulting in a decreasing amount of energy being deposited.

1967) finds that even at 5000 K, only 0.1% of the initial $N_2$ is dissociated to N atoms, but nevertheless 10% of the initial $N_2$ is incorporated into HCN. This is because of a Zel'dovich-like mechanism, in which a series of reactions between $N_2^*$ and various hydrocarbons and their radicals drive HCN formation for energies well below those required to dissociate $N_2$ alone. These include reactions such as $CH_2 + N_2^* \rightarrow CH + H+N_2$ (128 kcal mole$^{-1}$) and $CH + N_2^* \rightarrow HCN + N$. The activation energy for this last reaction has been determined in shock-tube experiments at temperatures between 2340 and 4660 K to lie between 14 kcal mole$^{-1}$ and 22 kcal mole$^{-1}$ (Dean et al. 1991; Lindackers et al. 1991; Medhurst et al. 1993). There is therefore no surprise that in $N_2/CH_4$ atmospheres, shock heating produces HCN at temperatures as low as 2500 K, and hydrocarbons at even lower temperatures. At temperatures above 5000 K, HCN production will increasingly be due to reactions with single N atoms, such as $CH_4 + N \rightarrow HCN + H_2 + H$ and analogous equations with $CH_4$ fragments (Ferus et al. 2017).

To calculate the total yields of HCN and hydrocarbons in Titan's atmosphere, we need production efficiencies $\phi_i$ (in molecules J$^{-1}$) for each of the $i$th species produced. At 4000 K, HCN production in Rao's shock-tube experiment corresponds to a production efficiency $\phi_{HCN} = 2.0 \times 10^{17}$ molecule J$^{-1}$ (Rao et al. 1967; Bar-Nun & Shaviv 1975). It is encouraging that this same production efficiency results from a thermochemical hydrodynamic model of shock heating due to a linear energy discharge based on Lin (1954)'s cylindrical shock modeling for an $N_2/CH_4$ atmosphere: for this theoretical model, HCN production efficiency $\phi_{HCN} = 1 \times 10^{17}$ molecules J$^{-1}$ for an assumed linear energy deposition of $E_0 = 10^5$ J m$^{-1}$ and an $N_2$/ $CH_4$ ratio of 4.3% $CH_4$ (Chameides & Walker 1981;

Borucki et al. 1988). The production efficiency depends only weakly and indirectly on $E_0$ with the variation due to changes in $T_F$ (which weakly depends on $E_0$). Chameides (1979) finds that the production efficiency (molecules J$^{-1}$) varies by a factor of less than three as $E_0$ ranges from 1 J m$^{-1}$ to $10^8$ J m$^{-1}$. But he suggests that values of $E_0$ below 1 J m$^{-1}$ produce temperatures too low for significant chemical yields, so we set this as our lower threshold for organic molecule production driven by shocks.

Figure 4 shows the energy deposition in J m$^{-1}$ for meteoroids with masses relevant to our discussion that enter, decelerate, and ablate in Titan's atmosphere. Meteoroids that enter with initial masses between 10 and $10^7$ g have maximum $E_0$ values ranging from 1 to $10^6$ J m$^{-1}$. Below, in evaluating Equation (11) we will take $R_1$ and $R_2$ to be the radii corresponding to spherical particles of these masses.

As previously noted, other investigators have used LIP experiments with $N_2/CH_4$ gas mixtures to simulate the effect of meteoroids and lightning (each of which produces a cylindrical shock wave) in Titan's atmosphere (Borucki & McKay 1987; Scattergood et al. 1989; Ferus et al. 2017). Borucki et al. (1988) and Scattergood et al. (1989) report results for a variety of $N_2/$ $CH_4/H_2$ concentrations. The closest of these to the observed Titan atmospheric conditions (95% $N_2$, 5% $CH_4$, 1 bar pressure) are those of Borucki et al. (1988). Scattergood et al. (1989) show that the results do not vary much for $H_2$ concentrations as high as 5%. Borucki & McKay (1987) report $\phi_i$ values for hydrogen cyanide (HCN), acetylene ($C_2H_2$), ethane ($C_2H_4$), ethylene ($C_2H_6$), and propane ($C_3H_8$), results that we have digitized from their published figures and summarized in Table 2. In Table 2, we compare the production efficiencies resulting from

**Table 2**
Comparison of Theoretical and Experimental Production Efficiencies $\phi_i$ (Molecules J$^{-1}$) for N$_2$/CH$_4$ Gas Mixtures

| Species | Thermochemical Calculation[a] | Shock Tube[b] | LIP[c] |
|---------|-------------------------------|---------------|--------|
| HCN | $1 \times 10^{17}$ | $2.0 \times 10^{17}$ | $6.0 \times 10^{16}$ |
| C$_2$H$_2$ | ... | $7.1 \times 10^{17}$ | $6.2 \times 10^{16}$ |
| C$_2$H$_4$ | ... | $1.0 \times 10^{17}$ | $5.2 \times 10^{15}$ |
| C$_2$H$_6$ | ... | ... | $4.5 \times 10^{15}$ |
| C$_3$H$_8$ | ... | ... | $1.3 \times 10^{14}$ |

**Notes.**
[a] Assumes $10^5$ J m$^{-1}$ and a 4.3% CH$_4$ atmosphere (Chameides & Walker 1981; Borucki et al. 1988).
[b] Results for 4000 K (Rao et al. 1967; Bar-Nun & Shaviv 1975).
[c] Assumes a 5% CH$_4$ atmosphere (Borucki et al. 1988).

thermochemical hydrodynamic calculations (Chameides & Walker 1981; Borucki et al. 1988) with those derived from shock-tube (Rao et al. 1967) and LIP experiments (Borucki et al. 1988). We note that as the temperature in the shock-tube experiments approaches 6000 K (results not displayed in Table 2 but tabulated by Bar-Nun & Shaviv 1975), the relative abundances of HCN, H$_2$H$_2$, and C$_2$H$_4$, and therefore their relative production efficiencies, approach those shown for the LIP experiments. But in fact the production efficiencies for HCN and are similar (within a factor of about three) across the shock-tube, LIP, and calculated results. LIP experiments achieve higher temperatures from the laser-induced shock than reached in most of the shock-tube experiments reported here, which likely accounts for the different relative $\phi_i$ reported in Table 4. Borucki & McKay (1987) and Scattergood et al. (1989) do not report the temperatures achieved in the LIP of their experiments, but Ferus et al. (2017) determine plasma temperatures in their LIP experiments ranging from 4200 to 9300 K.

## 5. Net Shock Synthesis in Titan's Atmosphere

Silber et al. (2017, 2018) estimate the temperatures $T'$ behind the cylindrical shock front of meteoroids entering Earth's atmosphere by relying on equations for strong shocks that relate the shock wave's Mach number $M$ to the pressure $p'$ behind the shock and the temperature $T_0$ and pressure $p_0$ of the unshocked atmosphere (Zeldovich & Raizer 2002):

$$\frac{p'}{p_0} = \frac{7}{6}M^2 - \frac{1}{6}, \tag{20}$$

and:

$$\frac{T'}{T_0} = \frac{1}{36}(7M^2 + 34 - 5M^{-2}). \tag{21}$$

Silber et al. (2017) note that Bronshten (1983) concludes that $10^2 \leqslant p'/p_0 \leqslant 10^4$ for meteoroids in Earth's atmosphere, and they choose $p'/p_0 = 100$. With this choice, Equation (20) gives the Mach number of the shock to be $M = 9.3$ and Equation (21) gives $T' = 17.6T_0$. By Equation (11), the majority of useful shock energy delivered to Titan's atmosphere will be due to the smallest particles capable of generating shocks with energies above 1 J m$^{-1}$, i.e., 10 g meteoroids. Figure 4 shows that 10 g meteoroids at Titan deposit the bulk of their energy around 400 km altitude. We see in Figure 1 that the temperature of Titan's undisturbed atmosphere at this altitude is about 180 K. Therefore for the Silber et al. (2017) choice of $p'/p_0 = 100$, the temperature behind the shock front reaches about 3200 K. This temperature is above the threshold for HCN and hydrocarbon generation found in shock-tube experiments. A choice $p'/p_0 = 200$, still near the very bottom of the range found by Bronshten (1983), would yield a temperature of 6600 K, in the realm of temperatures achieved in LIP experiments.

We also note that Equations (20) and (21) assume a specific heat ratio $\gamma = 1.4$ (Zeldovich & Raizer 2002) appropriate to a diatomic gas with two rotational degrees of freedom; including vibrational modes could lower $\gamma$, resulting in lower estimates for $T'$ (Zucrow & Hoffman 1976; Chameides 1979). We therefore consider the more general equations corresponding to Equations (20) and (21) but for arbitrary $\gamma$ (Anderson 2019). As implicit in Equations (20) and (21), we take the wave angle of the shock to be $\beta = \pi/2$, which should be correct for a cylindrical shock. We find:

$$\frac{p'}{p_0} = \frac{2\gamma}{\gamma + 1}M^2 - \frac{\gamma - 1}{\gamma + 1}, \tag{22}$$

and:

$$\frac{\rho'}{\rho_0} = \frac{(\gamma + 1)M^2}{(\gamma - 1)M^2 + 2}. \tag{23}$$

By the ideal gas law $T'/T_0 = (p'/p_0)/(\rho'/\rho_0)$, and we find:

$$\frac{T'}{T_0} = \frac{1}{(\gamma + 1)^2}[2\gamma(\gamma - 1)M^2 + 4\gamma - (\gamma - 1)^2 - 2(\gamma - 1)M^{-2}]. \tag{24}$$

It is easy to show that Equations (22) and (24) reduce to Equations (20) and (21) when $\gamma = 1.4 = 7/5$. Chameides (1979), relying on the discussion in Zucrow & Hoffman (1976), concludes that allowing $\gamma$ to range between 1.4 and 1.25 should adequately bound the effects due to the modes of energy transfer (such as the excitation of vibrational modes) beyond the rotational degrees of freedom. We there also calculate, as a lower end-member, $T'$ for $\gamma = 1.25 = 5/4$. We find by Equations (22) and (24) that $T' = 12.1T_0$ for $p'/p_0 = 100$, so at a 400 km altitude the temperature behind the shock front reaches about 2200 K, and about 4200 K for $p'/p_0 = 200$, a range that encompasses shock formation of both light hydrocarbons and HCN. The radius of the region in which the maximum meteoroid energy is deposited is the "characteristic radius" $R_0 = (E_0/p_0)^{1/2}$ (Silber et al. 2017, 2018). At a 400 km altitude, $p_0 = 1.4$ Pa, so $R_0$ varies between 0.8 m and 800 m as $E_0$ varies between 1 and $10^6$ J m$^{-1}$.

Of course no single experiment or calculation provides a perfect analog to shock-driven chemistry in Titan's atmosphere. Nevertheless, it is encouraging that the values for $\phi_{HCN}$ in Table 2 are as consistent as they are, falling within a factor $\sim$3 of one another. As previously noted, the majority of chemically useful shock energy delivered to Titan's atmosphere will be due to the smallest particles capable of generating shocks with energies above 1 J m$^{-1}$. Production efficiencies vary weakly with E$_0$, found in one calculation to drop by a factor of 2.3 as E$_0$ decreases from $10^5$ to 1 J m$^{-1}$ (Chameides 1979). To extrapolate from the $\phi_i$ values in Table 2 to those more appropriate for calculating the effects of these lower-energy cylindrical shocks, we therefore scale by a factor $f_\phi = 1/2.3 = 0.43$. It is likely that $f_\phi$ is species dependent, but to our knowledge this level of detail is not currently available either in theoretical or experimental results for $N_2/CH_4$ atmospheres. In our subsequent calculations, we will use the thermochemical hydrodynamic result in Table 2 for $\phi_{HCN}$, scaled down in magnitude by the factor $f_\phi$ to account for the fact that the bulk of the shock-synthesis-relevant energy delivered by meteoroids to Titan's atmosphere lies in objects with values of $E_0$ as low as ∼1 J m$^{-1}$. We then estimate the shock production efficiencies $\phi_{C2H2}$ and $\phi_{C2H4}$ from the shock-tube results in Table 2, scaled down by a factor 2 by analogy to the ratio of $\phi_{HCN}$ for the thermochemical hydrodynamic calculation versus the 4000 K shock-tube results, multiplying these values as well by $f_\phi$. These results, displayed in Table 3, provide the final production yield values we use in this study. We continue to carry two significant figures in Table 3 since these are intermediate values for our final calculations, but it should be clear from our discussion that these values are themselves reliable at best to one significant figure.

**Table 3**
Production Yield Values ($f_\phi \phi_i$) Used in This Study

| Species | $f_\phi \phi_i$ (molecules J$^{-1}$) |
|---|---|
| HCN | $4.3 \times 10^{16}$ |
| $C_2H_2$ | $1.5 \times 10^{17}$ |
| $C_2H_4$ | $2.2 \times 10^{15}$ |

With $v_i$ from Equation (3) and $M_q$ from Equation (11), the total energy flux going into Titan's atmosphere from meteoroids in the size range $R_1$–$R_2$ may be written as:

$$E_q = \frac{1}{2} M_q v_i^2, \qquad (25)$$

where $E_q$ has units J m$^{-2}$ s$^{-1}$. The number of molecules of species $i$ synthesized in units m$^{-2}$ s$^{-1}$ due to meteor shocks in Titan's atmosphere may then be estimated as:

$$N_{q,i} = f_\phi \phi_i E_q. \qquad (26)$$

By Equations (11) and (25) with $b + 3 = -0.643$, $E_q$, and so $N_{q,i}$, are dominated by the smallest meteoroid size $R_1$ that can generate a shock in the atmosphere. We take this to be the radius corresponding to a spherical particle of mass $m = 10$ g with $\delta = 3.4$ g cm$^{-3}$. Using $R_1 = 8.9$ mm in Equation (11) and $f_\phi \phi_i$ from Table 3, we calculate $N_{q,i}$ from Equation (26) for $q = 3$ and $q = 4$. We compare these with the production rates in Titan's atmosphere found via UV photochemistry (Krasnopolsky 2009) in Table 4. These column production rates are typically reported in the photochemistry literature in units of molecules cm$^{-2}$ s$^{-1}$ so we adopt these units for Table 4.

In Table 4 we also present the results for $H_2$ production. The photochemistry result is again from Krasnopolsky (2009). The results for shock production are not determined directly from experiments, but rather are estimated from the analysis of Civiš et al. (2017), for laser simulations of meteor shocks in $N_2/CH_4$ gases, with some mixtures also including $D_2O$. Absent of $D_2O$, the overall formation of acetylene from methane can be described by the reaction $2CH_4 \rightarrow C_2H_2 + 2H_2$, and of hydrogen cyanide by $2CH_4 + N_2 \rightarrow 2HCN + H_2$. Since these are the two dominant products from the shock chemistry (see Table 4), these reactions allow us to make a first estimate of $H_2$ production via shocks.

**Table 4**
Column Production Rates (cm$^{-2}$ s$^{-1}$) for HCN, Hydrocarbons, and $H_2$ in Titan's Atmosphere Produced by UV Photochemistry (from Krasnopolsky 2009, Table 6) Compared with the Meteoroid Shock Production Predicted Here Using the Tiscareno et al. (2013) Meteoroid Flux Models with $q = 3$ and $q = 4$

| Species | Photochemistry | Shocks ($q = 3$) | Shocks ($q = 4$) |
|---|---|---|---|
| HCN | $1.2 \times 10^9$ | $3 \times 10^7$ | $8 \times 10^4$ |
| $C_2H_2$ | $7.5 \times 10^9$ | $1 \times 10^8$ | $3 \times 10^5$ |
| $C_2H_4$ | $2.7 \times 10^9$ | $2 \times 10^6$ | $4 \times 10^3$ |
| $H_2$ | $1.2 \times 10^{10}$ | $2 \times 10^8$ | $6 \times 10^5$ |

Table 4 shows that for the case where $q = 3$ correctly models the ejecta distribution at Saturn's rings, meteor entry into Titan's atmosphere is responsible for ∼1% as much production of HCN, $C_2H_2$, and $H_2$ as is photochemistry. These percentages are reduced by a factor of about 300 if $q = 4$ provides a better ejecta model. Given how carefully tuned photochemistry models are now compared with the observed compositions for Titan's atmosphere (Hörst et al. 2018), these results imply that detailed Titan atmospheric organic chemistry models must henceforth take into account meteor shock production. This conclusion is only reinforced by a consideration of the altitude distribution of meteor shock synthesis.

## 6. Energy Deposition with Altitude

We showed in the previous section that meteoroid-shock-driven chemical synthesis likely makes a significant contribution to the production of organics and $H_2$ in Titan's atmosphere. We now describe the distribution of this production as a function of altitude.

First, we show that meteoric deposition of energy in Titan's atmosphere, and the resulting shock chemistry, have very different distributions than the energy inputted into Titan's atmosphere by the other larger sources of chemistry-driving exogenous energy incident upon Titan's atmosphere, UV photons and magnetospheric ions. We take the cumulative flux from the Tiscareno et al. (2013) models together with our model results of meteoroid deceleration and ablation through Titan's atmosphere to calculate the total atmospheric energy deposition in J km$^{-1}$ from meteoroids for one Saturnian year (29 Earth years). Using the photochemical model in Krasnopolsky (2009), we show in Figure 5 the total energy deposition of UV photons with wavelengths between 90–100 nm and 120–130 nm. Photons in these wavelength ranges are primarily responsible for $N_2$ and $CH_4$ photodissociation, respectively (see Figure 2 in Hörst et al. 2018). Smith et al. (2009) provide measurements of the $H^+$ flux onto Titan from the Cassini Magnetospheric Imaging Instrument (MIMI)'s Ion and Neutral Camera (INCA; Krimigis et al. 2004). Although there are many ion species in Saturn's magnetosphere, $H^+$ ions dominate in the outer portion through which Titan moves (Young et al. 2005). We present the energy input by $H^+$ in Titan's atmosphere in Figure 5 (Krimigis et al. 2004). In all cases, we determine the energy deposition by simply calculating the change in energy over altitude.

The quantitative results again depend on the correct choice of $q$ in the Tiscareno et al. (2013) ejecta model, but in the $q = 3$ model, we see that meteoric injection of energy exceeds that of magnetospheric protons and even that of 90–100 nm UV light. Regardless of the choice of $q$, meteoroids deposit their energy— and therefore drive atmospheric chemistry—hundreds of kilometers below the energy deposition by UV photons and protons, so that meteoroids appear to be the dominant source of chemistry-driving exogenous energy between about 250 and 600 km in altitude.

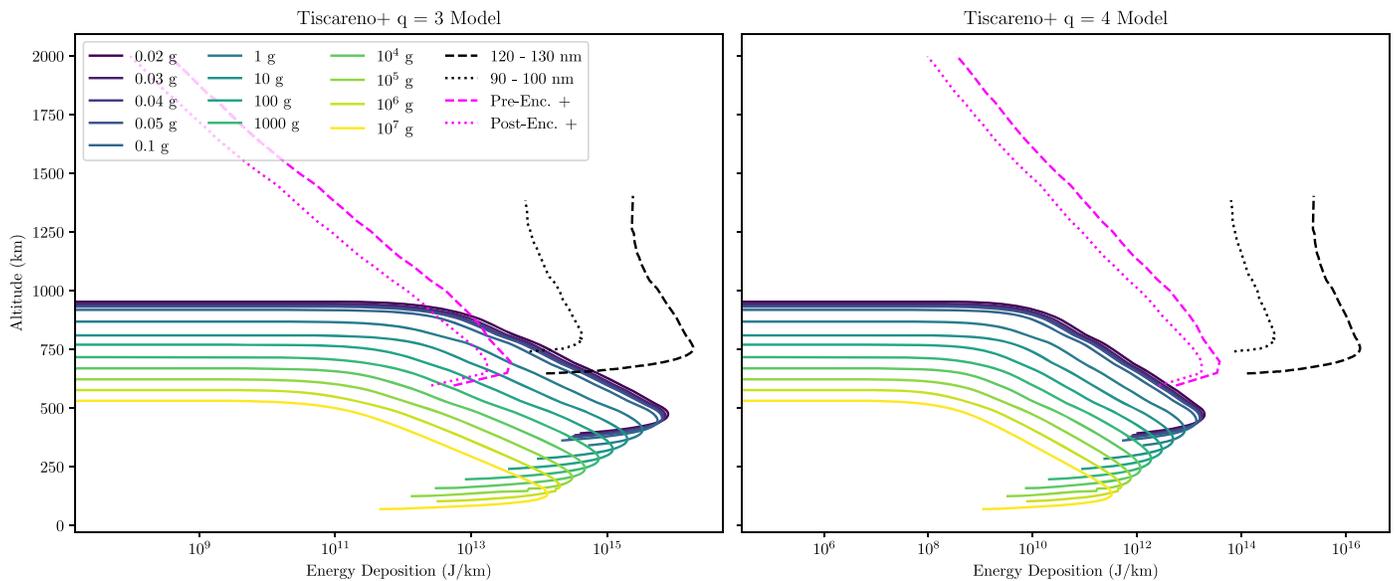

**Figure 5.** Comparison of the energy deposition in Titan's atmosphere due to all meteors hitting Titan's atmosphere in one Saturnian year (with flux distribution as described by the Tiscareno et al. 2013 model for $q = 3$ and $q = 4$; see text), compared to the energy deposition from magnetospheric H$^+$ measured by the Cassini MIMI INCA instrument pre- and postencounter with a Titan flyby (Krimigis et al. 2004; Smith et al. 2009), and the energy deposition from UV photons as modeled by Krasnopolsky (2009), also in one Saturnian year. The ion and photon curves invert at lower altitudes as the intensity of the ions/photons drops in the increasingly thick atmosphere.

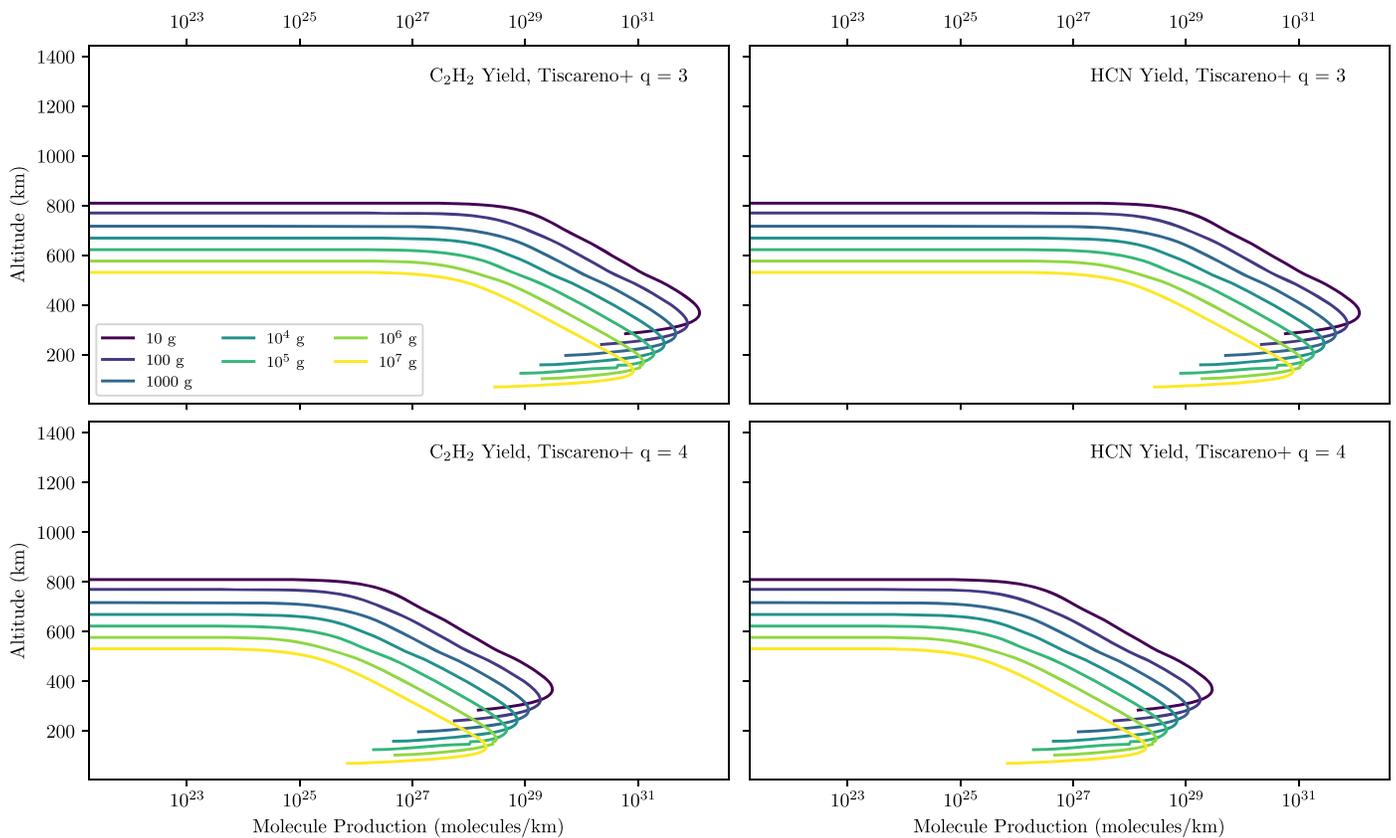

**Figure 6.** Number of molecules per kilometer produced by the energy generated from meteoroids shocking Titan's atmosphere for the $q = 3$ and $q = 4$ Tiscareno et al. (2013) flux cases, for one Saturnian year.

Since the efficiency of organic synthesis by shocks is often as high or higher than that for UV (Chyba & Sagan 1992), this result should hold for exogenously driven organic synthesis as well. In Figure 6, we show results from this simple model for the production of C$_2$H$_2$ and HCN per kilometer over a Saturnian year, by combining the results of Figure 5 with the efficiencies shown in Table 3. Between roughly 200 and 500 km in altitude, the production of HCN and $C_2H_2$ is dominated by meteoroid shock energy as UV photons and magnetospheric ions cannot penetrate the atmosphere as deeply. If $q = 3$, meteoroids deposit more energy than the magnetospheric $H^+$ ions and photons in some cases, while even for $q = 4$ they come within an order of magnitude of the $H^+$ energy deposition rate.

It is striking that this altitude broadly coincides with the presence of an observed haze layer (Sagan & Thompson 1984; Hörst 2017; Hörst et al. 2018). As seen in Figures 4 and 5, at the upper end of this range, meteoroids in the mass range 0.01–10 g fully ablate and thereby contribute their mass to this region of the atmosphere. This mass flux is given by Equation (11) with appropriate choices of $R_1$ and $R_2$. At the lower end of this range, the incoming meteoroids drive the production of organics via shock chemistry (Figure 6). In addition to the HCN and hydrocarbon production previously discussed, shock-tube experiments show that shocks in $N_2/CH_4$ atmospheres will also produce solid carbon that can then serve as nuclei for the formation of larger particles (Rao et al. 1967). At 4000 K, the production yield for solid carbon is almost twice that of the yield for HCN (Rao 1966; Rao et al. 1967). All this suggests a plausible connection between meteoroid bombardment of Titan's atmosphere and atmospheric haze, an investigation that we defer to a subsequent investigation.

## 7. Conclusion

In this study we have presented a model for meteoroid-driven organic chemistry in Titan's atmosphere, calculating the total resulting yields for hydrogen cyanide, acetylene, and ethane, as well as an implied resulting yield for molecular hydrogen. In order for an entering particle to shock Titan's atmosphere to a temperature sufficient to drive the relevant chemistry, the particle needs to satisfy two criteria.

First, it needs to be big enough so that the size of its vapor cap (much larger than the particle's physical diameter) is larger than an atmospheric molecule's mean free path—a comparison that of course depends on the altitude in the atmosphere (Silber et al. 2017, 2018). Our Knudsen number analysis and atmospheric entry model in Section 3, led us to conclude via Figure 3 that by this criterion shocks can be generated in Titan's atmosphere by meteoroids with masses $\geqslant 0.02$ g.

Second, we made use of our meteoroid atmospheric entry simulations to determine energy deposition (J m$^{-1}$) in the atmosphere over the course of the meteoroid's trajectory as that object decelerates and ablates. Previous work modeling the atmospheric chemistry that occurs as a result of linear shocks caused by meteoroid trajectories (Chameides 1979; Chameides & Walker 1981) concluded that a meteoroid would need to deposit at least 1 J m$^{-1}$ of energy, corresponding for meteoroid trajectories in Titan's atmosphere to masses above 10 g, to drive such chemistry.

It is now possible to estimate the flux of objects larger than 10 g entering Titan's atmosphere due to observations made by the Cassini spacecraft of apparent meteoroid strikes on Saturn's rings (Tiscareno et al. 2013). We used these results, albeit with a different choice of meteoroid density, to determine the net energy deposition in Titan's atmosphere (Section 2) and via our atmospheric entry simulations, the net energy deposition as a function of altitude (Section 6). The close agreement of thermochemical simulation results (Chameides & Walker 1981) and experimental shock-tube results (Rao 1966; Rao et al. 1967) for $N_2/CH_4$ atmospheres then allowed us to estimate

**Table 5**
Annual (Saturnian Year) Mass (kg) Production on Titan Due to Meteoroid Shocks Predicted Here Using the Tiscareno et al. (2013) Meteoroid Flux Models with $q = 3$ and $q = 4$

| Species | Shocks ($q = 3$) | Shocks ($q = 4$) |
|---|---|---|
| HCN | $1 \times 10^9$ | $3 \times 10^6$ |
| $C_2H_2$ | $4 \times 10^9$ | $1 \times 10^7$ |
| $C_2H_4$ | $7 \times 10^7$ | $1 \times 10^5$ |
| $H_2$ | $5 \times 10^8$ | $2 \times 10^6$ |

HCN, $C_2H_2$, $C_2H_4$, and $H_2$ production as a function of altitude in Titan's atmosphere (Sections 4, 5 and 6). We found that meteoroid-driven production of these molecules appears to be the dominant source for these molecules between roughly 200 and 500 km in Titan's atmosphere, an entirely new result. We suggested that may be a connection between this result and the atmospheric haze observed at this altitude (Sagan & Thompson 1984; Hörst 2017). Net production by meteoroid shocks in the atmosphere could reach as high as ∼1% of the photochemical production for these molecules, although that result is dependent on one's choice of the fragmentation parameter $q$ in Tiscareno et al.'s models for observations of impacts on Saturn's rings (Tiscareno et al. 2013). In Table 5 we use the results in Table 4 to provide the annual (Saturnian year, or 29.5 Earth years) production rates (kg yr$^{-1}$) for HCN, $C_2H_2$, $C_2H_4$, and $H_2$ on Titan.

Our model is simple in several ways. We have used average meteoroid velocities, ignoring the impact-velocity asymmetry between the leading and trailing faces of Titan (Section 2). We have taken the $N_2/CH_4$ ratio of Titan's atmosphere to be constant with altitude, but of course this is an approximation (e.g., see Niemann et al. 2005). And while we have used the thermochemical and shock-tube data available in the literature, and noted that these seem appropriate in relation to LIP experimental results, one can certainly wish that there were a more extensive set of both modeling results and experimental data exploring a wider range of atmospheric compositions, linear energy depositions, and shock temperatures. In future work we hope to relax at least some of these assumptions, while recognizing that considerable uncertainty will remain until there are additional spacecraft data addressing the flux of meteoroids more massive than 10 g in the vicinity of Saturn.

### Acknowledgments

This material is based upon work supported by the National Science Foundation Graduate Research Fellowship under grant No. DGE-1656466. Any opinions, findings, and conclusions or recommendations expressed in this material are those of the author(s) and do not necessarily reflect the views of the National Science Foundation.

This work has made use of data from the NASA, the European Space Agency (ESA), and Italian Space Agency (ASI) *Cassini-Huygens* mission (http://solarsystem.nasa.gov/missions/cassini/overview/).

We would like to thank M. Tiscareno for his insight and his patience with answering our questions on his 2013 study. We also would like to thank P.J. Thomas for his discussions about atmospheric models.

## ORCID iDs

Erin E. Flowers  https://orcid.org/0000-0001-8045-1765
Christopher F. Chyba  https://orcid.org/0000-0002-6757-4522